# Experimental Evidence of Guided Resonances in Photonic Crystals with Aperiodically-Ordered Supercells

Armando Ricciardi,[1,4] Marco Pisco,[2] Ilaria Gallina,[3] Stefania Campopiano,[1] Vincenzo Galdi,[3] Liam O' Faolain,[4] Thomas F. Krauss [4] and Andrea Cusano[2,*]

[1] Department for Technologies, University of Naples "Parthenope," I-80143, Naples, Italy

[2] Optoelectronic Division, Department of Engineering, University of Sannio, I-82100 Benevento, Italy

[3] Waves Group, Department of Engineering, University of Sannio, I-82100 Benevento, Italy

[4] School of Physics and Astronomy, University of St Andrews, St Andrews, Fife, KY16 9SS, UK

* Corresponding author: acusano@unisannio.it

ABSTRACT: We report on the first experimental evidence of *guided resonances* (GRs) in photonic crystal slabs based on *aperiodically-ordered* supercells. Using the Ammann-Beenker (quasiperiodic, 8-fold symmetric) tiling geometry, we present our study on the fabrication, experimental characterization, and *full-wave* numerical simulation of two representative structures (with different filling parameters) operating at near-infrared wavelengths (1300-1600 nm). Our results show a fairly good agreement between measurements and numerical predictions, and pave the way for the development of new strategies (based, e.g., on the lattice symmetry breaking) for GR engineering.

The reflection/transmission spectra of planar dielectric layer diffraction gratings can exhibit well-pronounced Fano-like sharp resonances attributable to the coupling of the externally propagating diffracted orders with the modes supported by the equivalent waveguide (i.e., the grating) [1]. Fan *et al*. [2], who analyzed these resonant *leaky* modes in photonic crystal (PC) slabs for the first time, defined them as "guided resonances" (GRs). Such resonances have recently attracted considerable research interest in view of the intriguing possibilities of tailoring/engineering their spectral features (e.g., by acting on the incidence angle and on the physical and geometrical parameters of the gratings) as well as their relevance for a number of applications. More specifically, GRs in PC slabs have been exploited to demonstrate mirrors [3], optical filters [4], polarization splitters [5] and displacement sensors [6], just to mention a few. The observation of GRs in asymmetric PC slabs has been also reported [7], as well as the characterization of their dependence on the illumination angle [8]. However, it is worth noting that all of these experiments, as well as most of the numerical work, have focused only on *periodic* PCs.

In recent work, Prasad *et al*. [9] argued that *strict* periodicity is an essential requirement for the excitation of GRs. However, in some recent studies [10-12], we have already demonstrated by numerical analysis that GRs can be excited in PC slabs based on *aperiodically ordered* supercells, and have also shown the possibility of controlling the GR excitation by introducing *ad-hoc* point defects [12]. In spite of these first numerical results, to the best of our knowledge, the excitation of GRs in *globally aperiodic* photonic quasicrystals (PQC) or in PC slabs with aperiodically-ordered supercells has never been experimentally observed. Starting from the periodic-approximant case, for which rigorous *full-wave* numerical modeling [10-12] is still computationally affordable, we now report the first experimental evidence of GR excitation in PCs with aperiodically-ordered supercells.

In accordance with our previous papers [10-12], we considered a supercell based on a *quasiperiodic* (Ammann-Beenker, octagonal) lattice exhibiting eight-fold symmetry [13] and containing 97 elements (or a fraction of them), shown in Fig. 1a.

We chose a silicon-on-insulator (SOI) wafer characterized by a top silicon layer with a nominal thickness $t$=220nm, 2 µm of buried oxide, and about 750 µm of bulk silicon for the experimental realization. The PQC parameters were chosen on the basis of the design carried out in [10], but taking fabrication constraints into account, e.g., leaving the buried oxide layer underneath the guiding layer for mechanical support and making holes with a radius larger than 90 nm in order to guarantee that they are properly etched into the silicon slab. Based on these considerations, as a first attempt for our study, we chose a lattice constant (see Fig. 1a) $a$=300 nm and a hole radius $r$=95 nm, corresponding to a filling parameter $r/a$= 0.32. Since the focus of this first study is on providing some generic experimental evidence of the GR phenomenon in PQCs, the design parameters were not optimized for a specific application.

A schematic of the main steps of the fabrication process is illustrated in Fig.1b. The pattern was first defined in ZEP-520A electron beam resist (by using a ZEISS GEMINI 1530/RAITH ELPHY PLUS electron beam writer) and then transferred into the top silicon layer by using low power, low DC bias, reactive ion etching (RIE) with a combination of $SF_6$ and $CHF_3$ gas (50:50 mix). At the end of the fabrication process, the remaining resist was removed by soaking in dimethlyacetamide. Figure 2 shows scanning electron microscope (SEM) images of a fabricated structure. The size of the patterned area was about 300 µm × 300 µm, corresponding to a periodic replication of about 14700 supercells (of size ~ 2.5 µm × 2.5 µm ).

The fabricated sample was then characterized in reflection using normal incidence illumination through a beamsplitter with the electric field polarized along the side of the supercell. We used a broadband supercontinuum (KOHERAS Super Compact) source. The spot size was about 60 µm (covering an area corresponding to nearly 600 supercells) We normalized the measured reflectivity of the slab to that of a gold mirror. Figure 3 shows the normalized reflectivity (measured with a resolution of 1 nm) within the wavelength range 1300-1600nm (chosen in view of the experimental limitations), and compares it with the numerical prediction obtained by finite-element modeling (COMSOL Multiphysics, RF Module [14]). The structural parameters used for the simulation, which closely represented those of the experimental realization as obtained by SEM inspection, are reported in the figure caption. The material dispersion is taken into account via a cubic spline interpolation of the tabulated values for silicon and silica [15]. The computational domain (restricted, for symmetry considerations, to one quarter of the supercell; see the dashed square in Fig. 1a) is transversely terminated with two horizontal perfectly-electric-conducting (PEC) and two vertical perfectly-magnetic-conducting (PMC) walls, so as to simulate a normally-incident plane-wave with vertically-polarized electric field. Air and silicon layers with a thickness of $4t$ are assumed at the top of the silicon slab and the bottom of the silica layer, respectively, along the *out-of-plane* axis. The entire structure is discretized using a hexahedral mesh of at least 15 lines per wavelength (resulting in about $2 \cdot 10^5$ unknowns).

The first result is shown in Fig. 3, where we observe that the classical Fabry-Perot behavior of the SOI slab is broken by a set of Fano-type resonances arising from the coupling of the incident plane wave with degenerate modes of the PC slab [10]. The positions of the three measured resonances (1333, 1472, and 1592 nm) are in good agreement with their numerical counterparts (1340, 1473, and 1588 nm, respectively). This important evidence indicates that full-wave design tools already successfully employed for standard periodic structures can still be effectively used in the more complex aperiodically-ordered case to design and tailor the GR properties (e.g., number and spectral positions) for specific applications.

In order to strengthen the aforementioned experimental evidence, we modified the slab design so as to improve the visibility of the GRs in the reflectivity spectrum. To this aim, we fabricated and characterized another structure with the same lattice constant ($a$=300 nm) but with larger holes ($r$=115 nm), i.e., $r/a$=0.38. In Fig. 4, the measured reflectivity response is compared with the numerical prediction. As predicted, this new structure ensures a higher visibility of the GR occurring in the investigated wavelength range. More specifically, it exhibits a deep broadband resonance at 1373 nm. The small shift compared to the numerically modeled resonance (found at 1398 nm) is due to fabrication-dependent variations in the hole diameter (typically $\pm$ 5%, $\pm$10nm) [16] and the variation in thickness of the SOITEC wafer (typically $\pm$ 5 nm).

As evident from the resemblance of the field patterns compared in Fig. 5, this new broadband resonance corresponds to that observed at around 1588 nm in the previous structure (with $r/a$=0.32, cf. Fig. 3), confirming that increasing the hole radius causes a blue shift of the resonances.

Moreover, as expected, the resonance bandwidth (about 30 nm, in good agreement with the numerical prediction) significantly increases; this can be explained by considering that increasing the hole size (i.e., reducing the "average" refractive index of the slab from ~2.5 for $r/a$=0.32 to ~2 for $r/a$=0.38) results in a shorter life-time of the GR, thus leading to a reduction of the quality factor. Although, at this stage, no effort was made to obtain GR-based cavities characterized by high quality factors, it is expected that significant optimization margins exist by acting on the tiling geometry, as well as on the filling factor, structure symmetry, slab thickness, and refractive index contrast.

An important aspect that emerges from the experimental results is the *robustness* of the GR phenomena in spite of the increased complexity of the slab supercell, taking also into account the related fabrication tolerances. This consideration, which cannot been taken for granted, paves the way to the exploration of GR effects in more complex structures, such as point-defected aperiodic supercells that enable the excitation of (otherwise uncoupled) GRs by controlling the symmetry properties of the spatial arrangement of the holes [12]. In fact, for this particular application, aperiodically-ordered lattice geometries seem to be inherently suited in view of the variety of inequivalent defect sites that they can offer by comparison with periodic PCs. Moreover, we have shown in some preliminary studies [17] that aperiodically-ordered PCs can also provide new degrees of freedom in tailoring the GR electromagnetic field distributions, which may strongly influence the tuning and sensitivity efficiency of GR-based photonic devices, thereby opening up interesting application perspectives.

In conclusion, we have experimentally demonstrated the evidence of GRs in PC slabs based on aperiodically-ordered supercells. A good agreement between numerical and experimental results has been observed. These results, which confirm our previous numerical predictions [10,11], may open up new intriguing perspectives on the engineering of GRs. In this framework, current and future investigations are aimed at the exploitation of the above effects in tuning/sensing scenarios [17], as well as the fabrication and experimental characterization of *free-standing* structures. Also of great conceptual interest, in order to further address basic questions concerning the necessity of strict periodicity, is the experimental verification of the GR phenomenon in *globally* aperiodic (i.e., not periodically replicated) structures [13], for which numerical studies would be computationally unaffordable.

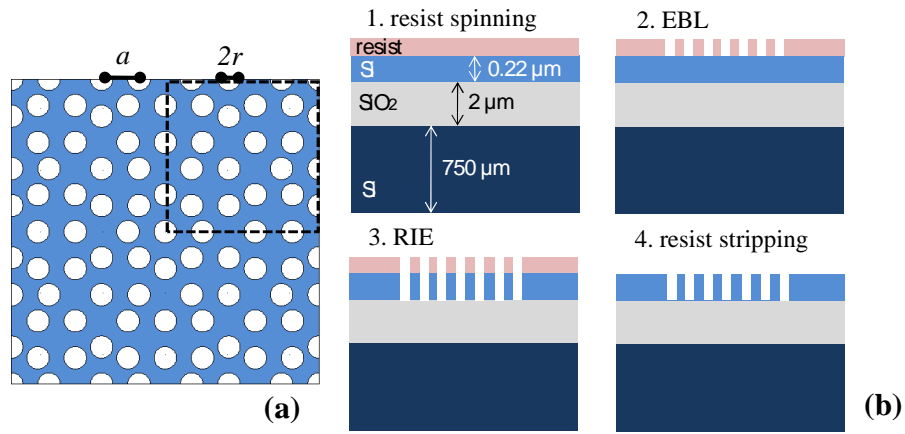

Fig.1. (Color online) (a) Ammann-Beenker (octagonal) supercell, with the lattice constant *a* and the hole radius *r* indicated. (b) Schematic of the fabrication process (details in the text).

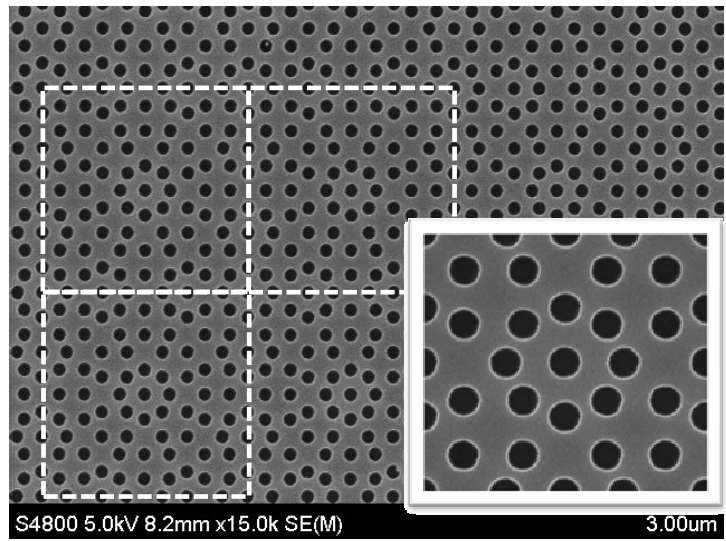

Fig. 2. SEM image (zoomed in the inset) of the fabricated PC slab with lattice constant *a*=300 nm and hole radius *r*=95 nm (i.e., *r*/*a*=0.32), with the white-dashed lines delimiting the supercells.

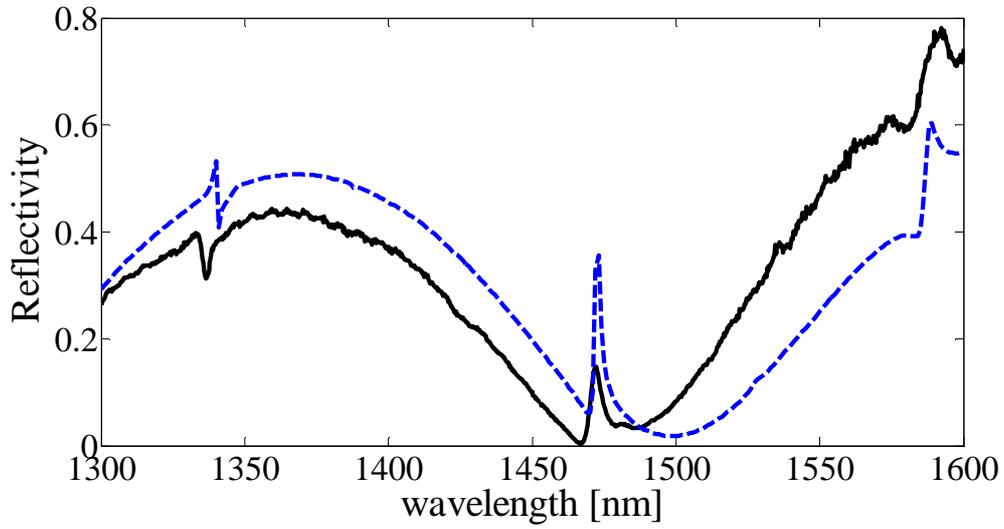

Fig. 3. (Color online) Measured (black-solid) and simulated (blue-dashed) reflectivity spectra for a PC slab with $a$=300 nm, $r$ =95 nm (i.e., $r/a$=0.32), and $t$=215 nm.

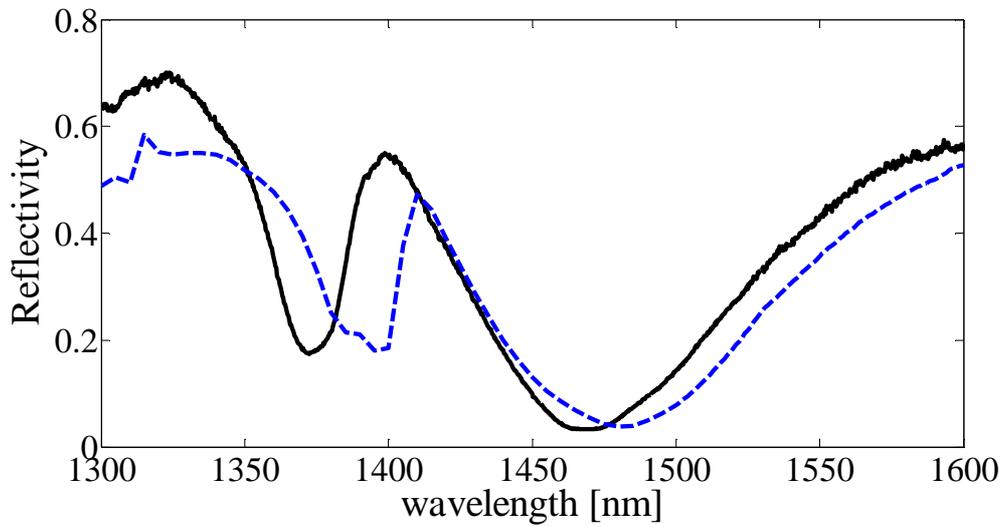

Fig. 4. (Color online) As in Fig. 3, but for a PC slab with $a$=300 nm, $r$ =115 nm (i.e., $r/a$= 0.38), and $t$=215 nm.

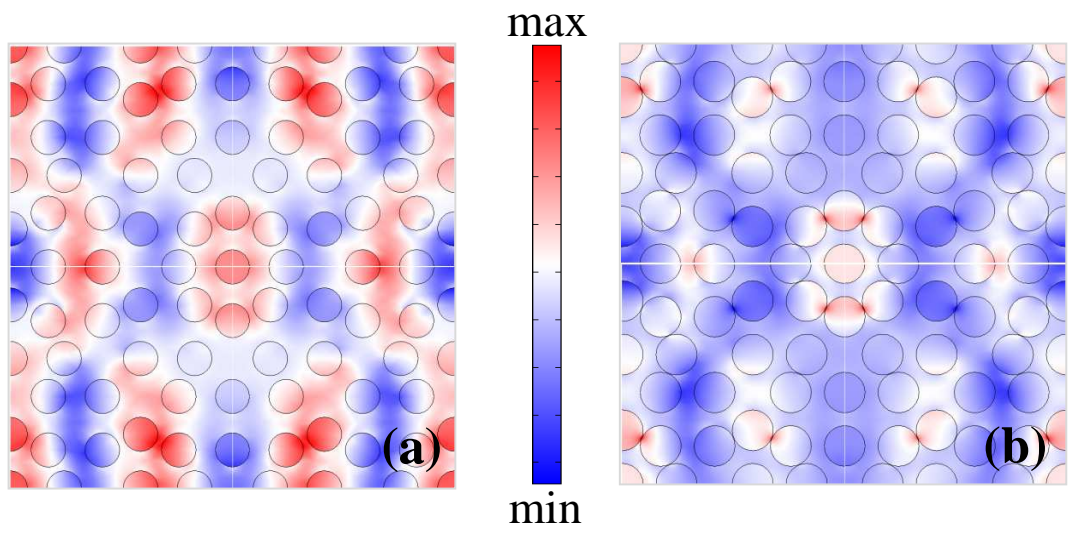

Fig. 5. (Color online) Simulated maps (in the transverse plane at the center of the slab) of the resonant electric fields for the PC slabs with $r/a$=0.32 at 1588 nm (a), and $r/a$=0.38 at 1398nm (b).